\documentclass[superscriptaddress,preprint]{revtex4}
\usepackage{amssymb}
\usepackage[tbtags]{amsmath}
\usepackage{graphicx}
\usepackage{epsfig,graphicx,times}
\usepackage{longtable}

\begin{document}

\title{Simplified approach to generate controlled-NOT gates with single trapped ions for arbitrary Lamb-Dicke parameters}
\author{Miao Zhang\footnote{Corresponding author: zhangmiao079021@163.com}, H.Y. Jia and L.F. Wei}
\affiliation{Laboratory of Quantum Opt-electronic Information,
Southwest Jiaotong University, Chengdu 610031, China}
\date{\today}

\begin{abstract}
For certain {\it specific} (or {\it``magic"}) Lamb-Dicke (LD)
parameters, Monroe {\it et al} showed [Phys. Rev. {\bf A 55}, R2489
(1997)] that a two-qubit quantum operation, between the external and
internal degrees of freedom of a single trapped ion, could be
implemented by applying a single carrier laser pulse. Here, we
further show that, such a two-qubit operation (which is equivalent
to the standard CNOT gate, only apart from certain phase factors)
could also be significantly-well realized for {\it arbitrarily}
selected LD parameters. Instead of the so-called ``$\pi$-pulses"
used in the previous demonstrations, the durations of the pulses
applied in the present proposal are required to be accurately set
within the decoherence times of the ion.
We also propose a simple approach by using only one off-resonant
(e.g., blue-sideband) laser pulse to eliminate the unwanted phase
factors existed in the above two-qubit operations for generating the
standard CNOT gates.



\end{abstract}

\maketitle

It has been shown that a quantum computer can be built using a
series of one-qubit operations and two-qubit controlled-NOT gates,
because any computation can be decomposed into a sequence of these
basic logic operations~\cite{fundamentalCNOT1}. Therefore,
precondition work is to effectively implement these fundamental
logic gates~\cite{fundamentalgates}. Since the first idea, proposed
by Criac and Zoller~\cite{Cirac95} in 1995 for implementing quantum
computation with trapped cold ions, much attention has been paid to
implement the fundamental quantum logic gates in the systems of
trapped cold
ions~\cite{Monroe95,Exprimental-research1,Two-ion-CNOT,8ions}.
Actually, a single CNOT logic operation  between the external and
internal states of a single trapped had been experimentally
demonstrated with $^{9}\text{Be}^{+}$ ion in 1995~\cite{Monroe95}.
Later, the CNOT gate between two individual trapped ions (i.e.,
$^{40}\text{Ca}^{+}$ ions) had also be experimentally implemented in
2003~\cite{Two-ion-CNOT}. Recently, quantum manipulations on eight
trapped ions had already been realized~\cite{8ions}.

However, most of these demonstrated experiments are operated within
the usual LD limit (wherein the spatial dimension of the ground
state of the collective motion of the ions is required to be much
smaller than the effective wavelength of the applied laser wave.)
i.e., the so-called LD parameters should be sufficiently small (see,
e.g.,~\cite{Monroe95}). In principle, quantum motion of a single
trapped ion beyond the above LD limit is also
possible~\cite{Weistates}. Furthermore, it has been shown that
utilizing the laser-ion interaction outside the LD regime might be
helpful to reduce the noise in the trap and improve the cooling rate
of the ion~\cite{Stevens1997}. Indeed, several approaches have been
proposed to coherently operate trapped ions beyond the LD limit for
implementing the desirable quantum logic gates~\cite{Weigates}.

More interestingly, beyond the LD limit Monroe {\it et
al}~\cite{monroe97} had shown that two-qubit quantum gates could be
implemented by using only one laser pulse. This is significantly
different from the previous scheme demonstrated within LD regime,
wherein three steps laser pulses are usually required. Although it
is relatively simple, approach in Ref.~\cite{monroe97} only work
well for certain {\it specific} LD parameters. However, accurately
setting the desirable LD parameter is not easy for the experiments.
In this Brief Report, given the LD parameter is arbitrarily set we
show that the two-qubit operations proposed in Ref.~\cite{monroe97}
could still be implemented sufficiently well. Furthermore, by adding
only one off-resonant (e.g., blue-sideband) we propose a simple way
pulse to eliminate the unwanted phase factors existed in the above
two-qubit operation for generating the standard CNOT gates. This
means that, for arbitrary LD parameters the exact single-ion CNOT
gate could be sufficiently-well implemented by using two laser
pulses. Besides the requirement of an auxiliary atomic level, the
present proposal for implementing the standard CNOT gate is really
simpler than many previous ones, including that proposed recently in
Ref.~\cite{Our-PRA}. Here, the so-called standard CNOT gate between
the external and internal degrees of freedom of the ion
reads~\cite{Monroe95}
\begin{equation}
\hat{C}_N=|0\rangle|g\rangle\langle0|\langle
g|+|0\rangle|e\rangle\langle0|\langle
e|+|1\rangle|g\rangle\langle1|\langle
e|+|1\rangle|e\rangle\langle1|\langle g|,
\end{equation}
with $|g\rangle$ and $|e\rangle$ being two selected internal atomic
levels and $|0\rangle$ and $|1\rangle$ the two lowest motional Fock
states of the ion's external vibration.

We consider that a single ion is trapped in a coaxial resonator RF
(radio frequency)-ion trap~\cite{BA00,N00}, and assume that only the
quantized vibrational motion along the principal $x$ axe is
important for the cooled ion. Following Monroe {\it et
al.}~\cite{monroe97} the ion is driven by a classical traveling-wave
laser field (with frequency $\omega_L$ and initial phase
$\theta_L$). In the rotating framework (rotating with the angular
frequency $\omega_L$), the system can be described by the following
Hamiltonian~\cite{monroe97,Vogel95}
\begin{eqnarray}
%\begin{array}{l}
\hat{H}=\hbar\nu(\hat{a}^\dagger
\hat{a}+\frac{1}{2})+\frac{\hbar\delta}{2}\hat{\sigma}_z+\frac{\hbar\Omega}{2}
\{\hat{\sigma}_+\exp[i\eta(\hat{a}+\hat{a}^\dagger)-i\theta_L]+H.c.\}.
%\end{array}
\end{eqnarray}
Here, $\hat{a}^\dagger$ and $\hat{a}$ are the bosonic creation and
annihilation operators of the external vibrational quanta (with
frequency $\nu$) of the ion. The Pauli operators
$\hat{\sigma}_z=|e\rangle\langle e|-|g\rangle\langle g|$ and
$\hat{\sigma}_+=|e\rangle\langle g|$ are defined by the internal
ground state $|g\rangle$ and excited state $|g\rangle$ of the ion,
respectively. The Rabi frequency $\Omega$ describes the coupling
strength between the ion and the applied laser beam. Also,
$\delta=\omega_0-\omega_L$ with $\omega_0$ being the eigenfrequency
of the target qubit generated by the two selected atomic levels
$|g\rangle$ and $|e\rangle$. Finally, $\eta$ is the LD parameter
describing the coupling strength between the atomic levels and
external vibrational quanta of the ion.

Suppose that the ion is driven by applying the laser to the $k$th
blue-sideband vibration, i.e., the frequency of the applied laser
beam is chosen as $\omega_L=\omega_0+k\nu$ with $k$ being a positive
integer. Without performing the LD approximation and under the usual
rotating-wave approximation, we have the following simplified
Hamiltonian~\cite{Weistates,Weigates}
\begin{eqnarray}
%\begin{array}{l}
\hat{H}_I=\frac{\hbar\Omega}{2}e^{-\eta^2/2}\big[(i\eta)^ke^{-i\theta_L}
\hat{\sigma}_+\sum_{j=0}^{\infty}\frac{(i\eta)^{2j}}{j!(j+k)!}(\hat{a}^{\dagger})^{j+k}\hat{a}^{j}+H.c.\big]
%\end{array}
\end{eqnarray}
in the interaction picture defined by the unitary operator
$\hat{U}_0=\exp\{-it[\nu(\hat{a}^\dagger\hat{a}+1/2)+\delta\hat{\sigma}_z/2]\}$.
The dynamics defined by this Hamiltonian is exactly
solvable~\cite{Weigates}, and its corresponding dynamical evolutions
read:
\begin{eqnarray}
\left\{
\begin{array}{l}
|m\rangle|g\rangle\longrightarrow\cos(\Omega_{m,k}t)|m\rangle|g\rangle
+i^{k-1}e^{-i\theta_{L}}\sin(\Omega_{m,k}t)|m+k\rangle|e\rangle,\\

|m\rangle|e\rangle\longrightarrow|m\rangle|e\rangle,\,m<k,\\

|m\rangle|e\rangle\longrightarrow\cos(\Omega_{m-k,k}t)|m\rangle|e\rangle
-(-i)^{k-1}e^{i\theta_{L}}\sin(\Omega_{m-k,k}t)|m-k\rangle|g\rangle,\,m\geq
k,
\end{array}
\right.
\end{eqnarray}
with $|m\rangle$ being the number state of the external vibration of
the ion, and
\begin{equation}
\Omega_{m,k}=\frac{\Omega\eta^{k}}{2}\sqrt{\frac{(m+k)!}{m!}}e^{-\eta^{2}/2}\sum_{j=0}^{m}\frac{(i\eta)^{2j}m!}{j!(m-j)!(j+k)!}.
\end{equation}
being the effective Rabi frequencies. If $k=0$, i.e., a resonant
laser pulse (the ``carrier" pulse) of frequency
$\omega_{L}=\omega_{0}$ is applied to drive the trapped ion, then
the above exact dynamical evolutions take the time evolution
operator
\begin{equation}
\hat{C}(t_1,\theta_1)=
\begin{pmatrix} C_{11} & C_{12}e^{-i(\theta_1+\frac{\pi}{2})} & 0 & 0
\\ C_{21}e^{i(\theta_1-\frac{\pi}{2})} & C_{22} & 0 & 0
\\ 0 & 0 & C_{33} & C_{34}e^{-i(\theta_1+\frac{\pi}{2})}
\\ 0 & 0 & C_{43}e^{i(\theta_1-\frac{\pi}{2})} & C_{44}
\end{pmatrix},
\end{equation}
with
\begin{eqnarray}
\begin{array}{l}
C_{11}=\cos(\Omega_{0,0}t_1),\,\,\,\,\,\,\,\,
C_{12}=\sin(\Omega_{0,0}t_1),
\\
C_{21}=C_{12},\,\,\,\,\,\,\,\,\,\,\,\,\,\,\,\,\,\,\,\,\,\,\,\,\,\,\,
C_{22}=C_{11},
\\
C_{33}=\cos(\Omega_{1,0}t_1),\,\,\,\,\,\,\,\,
C_{34}=\sin(\Omega_{1,0}t_1),
\\
C_{43}=C_{34},\,\,\,\,\,\,\,\,\,\,\,\,\,\,\,\,\,\,\,\,\,\,\,\,\,\,\,
C_{44}=C_{33}.
\end{array}
\end{eqnarray}
in the subspace $\Gamma=\{|0\rangle|g\rangle, |0\rangle|e\rangle,
|1\rangle|g\rangle, |1\rangle|e\rangle \}$. Above, $\theta_1$ and
$t_1$ are the initial phase and duration of the applied carrier
laser pulse, respectively. They should be set up properly for
realizing the expected quantum logic operation within this subspace.

Obviously, if
\begin{equation}
c_{11}=c_{34}=1,
\end{equation}
then a two-qubit quantum operation~\cite{monroe97}
\begin{equation}
\hat{C}_1(\theta_1)=
\begin{pmatrix} 1 & 0 & 0 & 0
\\ 0 & 1 & 0 & 0
\\ 0 & 0 & 0 & e^{-i(\theta_1+\frac{\pi}{2})}
\\ 0 & 0 & e^{i(\theta_1-\frac{\pi}{2})} & 0
\end{pmatrix}
\end{equation}
could be implemented. This operation is equivalent to the standard
CNOT gate (1) between the external and internal states of the ion,
apart from the phase factors $\exp[-i(\theta_1+\pi/2)]$ and
$\exp[i(\theta_1-\pi/2)]$.

The above condition (8) could be satisfied by properly setting the
relevant experimental parameters: $t_1$ and $\eta$, as
\begin{eqnarray}
t_{1}=\frac{2n\pi}{\Omega_{0,0}}\,,
\,\eta^2=1-\frac{m-\frac{3}{4}}{n},\,\,\, n,m=1,2,3....,
\end{eqnarray}
with $n$ and $m$ being arbitrary positive integers. Note that in the
scheme of Monroe {\it et al}~\cite{monroe97}, a slightly different
condition (from Eq. (8)): $|c_{11}|=|c_{34}|=1$ is required. Under
such a condition the uncertain phase factors depend not only on the
initial phase $\theta_1$ but also on the LD parameters. This may
complicate the progress to eliminate the unwanted phase factors for
practically realizing the standard CNOT gate. Here, we begin with a
relatively simply condition (8).

Theoretically, condition (10) is always satisfied for
arbitrary-selected LD parameters by properly selecting the values of
the integers $n,m=1,2,3,...$. As a consequence, the two-qubit
operation (9) could be, in principle, implemented for arbitrary LD
parameters by properly setting the durations of the applied carrier
laser pulse. However, because the practical existence of
decoherence, as we discussed in~\cite{Our-PRA}, the duration of the
present pulse should be shorter than the decoherence times of both
the atomic and motional states of the
ion~\cite{Decoherence1,Decoherence2}. This limits that {\it the
integers $n$ could not take arbitrary large values to let Eq.~(10)
be exactly satisfied}. Experimentally, the lifetime of the atomic
excited states $|e\rangle$ reaches $1$ s~\cite{BA00,N00} and the
coherence superposition of $|0\rangle$ and $|1\rangle$ can be
maintained up to $1$ ms~\cite{Decoherence2}. For the robustness of
the experimental realization, we limit the decoherence time strictly
a little, e.g., $\lesssim 0.1$ ms for the experimental Rabi
frequency $\Omega/2\pi\approx500$ KHz~\cite{Wineland96}. Based on
these data we can always find, via numerical method, sufficiently
well approximated solutions to Eq. (8) for implementing the quantum
operation (9) with sufficiently high fidelities.

\begin{longtable}{ccccccc}
\caption{Numerical results for implementing quantum operation (9)
for arbitrary selected LD parameters; from $0.18$ to $0.98$ (step by
0.02). The duration of the applied carrier pulse is given by $\Omega
t_1$}
\endfirsthead
\endhead
\multicolumn{4}{}{}\\
\hline \hline $\eta$ \,& $t_{1}\Omega$ \,& $C_{11}=C_{22}$ \,&
$C_{12}=C_{21}$ \,& $C_{33}=C_{44}$ \,& $C_{34}=C_{43}$
\\
\hline
0.18 \,& 267.75 \,& 0.97520 \,& -0.22135 \,& -0.21954 \,& 0.97560 \\
0.20 \,& 243.65 \,& 0.99948 \,& 0.03218 \,& 0.03193 \,& 0.99949 \\
0.22 \,& 179.76 \,& 0.97284 \,& -0.23146 \,& -0.23053 \,& 0.97306 \\
0.24 \,& 168.13 \,& 1.00000 \,& 0.00000 \,& 0.00000 \,& 1.00000 \\
0.26 \,& 129.49 \,& 0.97165 \,& -0.23640 \,& -0.24006 \,& 0.97076 \\
0.28 \,& 130.92 \,& 0.99376 \,& 0.11153 \,& 0.11041 \,& 0.99389 \\
0.30 \,& 104.95 \,& 0.99505 \,& -0.09935 \,& -0.09778 \,& 0.99521 \\
0.32 \,& 92.35 \,& 0.99374 \,& -0.11172 \,& -0.10790 \,& 0.99416 \\
0.34 \,& 79.49 \,& 0.98271 \,& -0.18517 \,& -0.18852 \,& 0.98207 \\
0.36 \,& 80.64 \,& 0.99586 \,& 0.09088 \,& 0.09407 \,& 0.99557 \\
0.38 \,& 67.33 \,& 0.99541 \,& -0.09572 \,& -0.09377 \,& 0.99559 \\
0.40 \,& 68.44 \,& 0.98505 \,& 0.17225 \,& 0.16794 \,& 0.98580 \\
0.42 \,& 54.57 \,& 0.98859 \,& -0.15063 \,& -0.15383 \,& 0.98810 \\
0.44 \,& 55.56 \,& 0.99646 \,& 0.08411 \,& 0.08530 \,& 0.99636 \\
0.46 \,& 111.30 \,& 0.97957 \,& -0.20111 \,& -0.19843 \,& 0.98012 \\
0.48 \,& 41.83 \,& 0.97796 \,& -0.20881 \,& -0.20607 \,& 0.97854 \\
0.50 \,& 42.72 \,& 1.00000 \,& 0.00000 \,& 0.00000 \,& 1.00000 \\
0.52 \,& 43.67 \,& 0.97497 \,& 0.22234 \,& 0.21915 \,& 0.97570 \\
0.54 \,& 87.23 \,& 1.00000 \,& 0.00000 \,& 0.00000 \,& 1.00000 \\
0.56 \,& 132.93 \,& 0.96361 \,& 0.26731 \,& 0.26592 \,& 0.96400 \\
0.58 \,& 118.42 \,& 0.97544 \,& -0.22027 \,& -0.22025 \,& 0.97544 \\
0.60 \,& 29.81 \,& 0.99320 \,& -0.11640 \,& -0.11358 \,& 0.99353 \\
0.62 \,& 30.64 \,& 0.99720 \,& 0.07473 \,& 0.07201 \,& 0.99740 \\
0.64 \,& 31.52 \,& 0.96241 \,& 0.27158 \,& 0.26906 \,& 0.96312 \\
0.66 \,& 62.42 \,& 0.99952 \,& -0.03096 \,& -0.03027 \,& 0.99954 \\
0.68 \,& 95.26 \,& 0.99509 \,& 0.09898 \,& 0.09985 \,& 0.99500 \\
0.70 \,& 353.53 \,& 0.99224 \,& 0.12437 \,& 0.12458 \,& 0.99221 \\
0.72 \,& 178.61 \,& 0.97983 \,& -0.19981 \,& -0.20082 \,& 0.97963 \\
0.74 \,& 82.49 \,& 0.99876 \,& -0.04974 \,& -0.05286 \,& 0.99860 \\
0.76 \,& 50.12 \,& 0.99715 \,& -0.07548 \,& -0.07608 \,& 0.99710 \\
0.78 \,& 186.67 \,& 0.96719 \,& -0.25406 \,& -0.25835 \,& 0.96605 \\
0.80 \,& 155.88 \,& 0.99888 \,& 0.04737 \,& 0.04576 \,& 0.99895 \\
0.82 \,& 175.54 \,& 0.99258 \,& -0.12162 \,& -0.12311 \,& 0.99239 \\
0.84 \,& 17.27 \,& 0.97693 \,& -0.21356 \,& -0.21395 \,& 0.97685 \\
0.86 \,& 18.04 \,& 0.99867 \,& -0.05149 \,& -0.05191 \,& 0.99865 \\
0.88 \,& 18.88 \,& 0.99205 \,& 0.12582 \,& 0.12453 \,& 0.99221 \\
0.90 \,& 19.79 \,& 0.95031 \,& 0.31129 \,& 0.31157 \,& 0.95022 \\
0.92 \,& 153.85 \,& 0.99324 \,& 0.11610 \,& 0.11507 \,& 0.99336 \\
0.94 \,& 39.40 \,& 0.99517 \,& 0.09817 \,& 0.09647 \,& 0.99534 \\
0.96 \,& 60.04 \,& 0.99627 \,& 0.08632 \,& 0.08611 \,& 0.99629 \\
0.98 \,& 122.12 \,& 0.99709 \,& 0.07617 \,& 0.07482 \,& 0.99720 \\
\hline \hline
\end{longtable}

In table I we present some numerical results for setting proper
experimental parameters $\Omega t_1$, to implement quantum operation
(9) robustly for the arbitrarily selected LD parameters (not limited
within the LD regime requiring $\eta\ll1$) from $0.18$ to $0.98$.
It is seen that, the probability amplitudes $C_{11}=C_{22}$ and
$C_{34}=C_{43}$ are desirably large, most of them could reach to
$0.99$. While, unwanted probability amplitudes $C_{12}=C_{21}$ and
$C_{33}=C_{44}$ are really significantly small; all of them is less
than $0.32$. This implies that the lowest fidelity for implementing
the quantum operation (9) is larger than $90\%$.

Certainly, the above approximated solutions could be further
improved by either relaxing the limit from the decoherence time or
increasing Rabi frequency $\Omega$ (via increasing the powers of the
applied laser beams) to shorten the operational time. For example,
if the decoherence time of the external quantum vibration of the ion
(e.g., the superposition of the $|0\rangle$ and $|1\rangle$) is
relaxed to the experimentally measured value (i.e., $1$
ms)~\cite{Decoherence2}, then almost all the coefficients
$C_{11}=C_{22}$ and $C_{34}=C_{43}$ reach to about $0.999$ or more
larger. This implies that for arbitrarily LD parameters the
two-qubit gate (9) could always be realized for the ion with
sufficiently long decoherence time.
In principle, designing the applied laser pulse with so short
duration is not a great difficulty for the current experimental
technology, e.g., the femto-second ($10^{-15}$s) laser technique.
Also, our numerical calculations show that the influence of the
practically-existing fluctuations of the applied durations is really
weak. For example, for the Rabi frequency $\Omega/2\pi\approx500$
kHz, the fluctuation $\delta t\approx 0.1\,\mu$s of the duration
lowers the desirable probability amplitudes, i.e., $C_{11}$ and
$C_{34}$ presented in table I, just about $5\%$. Thus, even consider
the imprecision of the durations, the amplitude of the desirable
elements, $C_{11}$ and $C_{34}$, are still sufficiently large, e.g.,
up to about 0.95. Therefore, the approach proposed here to implement
the desirable quantum operation (9) for arbitrary LD parameters
should be experimentally feasible.

Finally, we consider how to generate the standard CNOT gate (1) with
a single trapped ion from the quantum operation (9) produced above.
This could be achieved by just eliminating the unwanted phase
factors in (9) via introducing another off-resonant laser pulse.
Indeed, a first blue-sideband pulse (of frequency
$\omega_L=\omega_{ea}+\nu$ and initial phase $\theta_2$) induces the
following evolution
\begin{equation}
|1\rangle|e\rangle\longrightarrow\cos(\Omega_{0,1}t_2)|1\rangle|e\rangle-e^{i\theta_{2}}\sin(\Omega_{0,1}t_2)|0\rangle|a\rangle,
\end{equation}
but does not evolve the states $|0\rangle|g\rangle$,
$|1\rangle|g\rangle$ and $|0\rangle|e\rangle$. Above, $|a\rangle$ is
an auxiliary atomic level~\cite{Monroe95}, and $\omega_{ea}$ being
the transition frequency between it and the excited state
$|e\rangle$. Obviously, a ``$\pi$-pulse" defined by
$\Omega_{0,1}t_{2}=\pi$ generates a so-called controlled-Z logic
operation
\begin{equation}
\hat{C}_2=
\begin{pmatrix}
   1 & 0 & 0 & 0
\\ 0 & 1 & 0 & 0
\\ 0 & 0 & 1 & 0
\\ 0 & 0 & 0 & -1
\end{pmatrix}
\end{equation}
For the LD parameters from $0.18$ to $0.98$, and
$\Omega/2\pi\approx500$ kHz, the durations for this implementation
are numerically estimated as $3.3\times10^{-3}\sim 1.2\times10^{-2}$
ms.
Therefore, the standard CNOT gate (1) with a single trapped ion
could be really implemented by only two sequential operations
demonstrated above, i.e., $\hat{C}_N=\hat{C_{1}}(\pi/2)\hat{C}_2$.

In summary, we have rechecked the scheme of Monroe {\it et
al}~\cite{monroe97} for implementing a two-qubit quantum operation
with a single trapped ion by using only a single carrier
``$\pi$-pulse" laser beam. We found that, if the limit of definite
decoherence time is not considered, then such an approach works
really for arbitrarily selected LD parameters, not limits to the
so-called ``magic" values. Our numerical results indicated that, if
the durations of the applied carrier pulses are properly set (rather
than that in the so-called ``$\pi$-pulse"), then the above two-qubit
quantum operation could still be implemented within the definite
decoherence time for arbitrarily selected LD parameters.
Also, we have discussed the influence from the possible fluctuations
of the durations on the implementations of the quantum operation,
and shown that such a influence is really weak. In addition, by
using a single blue-sideband laser pulse we have shown that the
unwanted phase factors induced by the above carrier driving could be
eliminated. Therefore, a standard CNOT gate with a single trapped
ion could be practically implemented by using only two laser pulses;
one carrier pulse pluses one off-resonant one. Finally, we hope that
the numerical results, presented in table I, might be useful for the
future experiment.

%\vspace{2cm}

\end{document}